# Modelling the skew and smile of SPX and DAX index options using the Shifted Log-Normal and SABR stochastic models

Jan Kukliński and Doinita Negru
*Faculté des Hautes Etudes Commerciales (HEC)*
*Université de Lausanne, CH-1015 Lausanne, Suisse*

Pawel Pliszka
*SunTrust RH Capital Markets, Atlanta GA 30326, USA*

We discuss modelling of SPX and DAX index option prices using the Shifted Log-Normal (SLN) model, (also known as Displaced Diffusion), and the SABR model. We found out that for SPX options, an example of strongly skewed option prices, SLN can produce a quite accurate fit. Moreover, for both types of index options, the SLN model is giving a good fit of near-at-the-forward strikes. Such a near-at-the-money fit allows us to calculate precisely the skew parameter without involving directly the 3rd moment of the related probability distribution. Eventually, we can follow with a procedure in which the skew is calculated using the SLN model and further smile effects are added as a next iteration/perturbation. Furthermore, we point out that the SLN trajectories are exact solutions of the SABR model for $\rho = \pm 1$.

## I. BACHELIER, BLACK – SCHOLES, SHIFTED LOG – NORMAL AND SABR MODELS.

The mathematical concept of Brownian motion and its application in option pricing was developed for the first time by L. Bachelier [1] back in 1900. Later on, in 1973, Black-Scholes-Merton [2] reintroduced the concept of Brownian motion, proposing an option pricing model which assumed the stock prices to be log-normally distributed, given that their dynamics follow a geometric Brownian motion. Following the stock market crash of 1987, the inadequacy of lognormal distribution became obvious and the industry adopted volatility skew as a mean to correct the option pricing, while still retaining Black-Scholes parameterization. In the early nineties local-volatility models (Dupire model) and stochastic volatility models (Heston model) attempted a self-consistent probabilistic description that can reflect market behaviour. However, these models proved to be not efficient enough. The wide-spread use of the Black-Scholes (BS) parameterization leads to the concept of implied (BS) volatility that is lacking probabilistic interpretation. To this end, increasingly many practitioners are using the Bachelier (B) model (referred often to as Normal-Model) as a basis for parameterization. Despite the existence of many models capable of fitting observed option prices, a utility of a very simple model that can be used as a basic parameterization is of practical significance, especially in pricing and risk management of structured products involving other payoff functions.

The Shifted Log-Normal model, also known as Displaced Diffusion local volatility model [3], [4], represents an algebraic interpolation between the log-normal (Black-Scholes) and the normal model (Bachelier). The call price of the vanilla option under SLN writes as

$$C^{SLN} = e^{-r\Delta t}\left\{(F-K)N(d_2) + \frac{F}{q}\left[N(d_1) - N(d_2)\right]\right\} \quad (1)$$

$$d_{1/2} = \frac{1}{q\sigma_q\sqrt{\Delta t}}\ln\left(\frac{F}{F-q(F-K)}\right) \pm \frac{q\sigma_q\sqrt{\Delta t}}{2} \quad (2)$$

These prices can be easily derived from the explicit expression of the stochastic price trajectories

$$\tilde{S}_t = S_0 e^{r\Delta t}\left\{1 + \frac{1}{q}\left[e^{|q|\sigma_q \widetilde{W}_t - \frac{q^2\sigma_q^2}{2}t} - 1\right]\right\} \quad (3)$$

where $\widetilde{W}_t$ is the trajectory of a unit Wiener process.

In this context, the Shifted Log-Normal model can be characterized by two parameters: the models sigma/volatility parameter $\sigma_q$ and the skew parameter $q$. The advantage of the SLN model becomes straightforward: unlike the Bachelier model which implies a normal symmetric price density and the Black-Scholes model with a positively skewed probability distribution function, SLN can produce negatively skewed distributions. As we will show later on, this feature is highly appreciable taking into consideration that the implied volatility surfaces of SPX and DAX index options exhibit negatively skewed patterns.

The Skew and Smile are related to two properties of the distorted normal distribution: skewness and kurtosis. Skewness is defined as $\gamma = \mu_3/\sigma^3$ where $\mu_3$ is the third central moment of a probability distribution. Consequently, Kurtosis is presented as $k = (\mu_4/\sigma^4 - 3)$. Practical option pricing leads to loosely defined skew and smile which are qualitatively related to skewness and kurtosis. An important feature of the SLN model is that the skewness has a simple algebraic form

$$|\gamma| = \left(e^{q^2\sigma_q^2} + 2\right)\sqrt{e^{q^2\sigma_q^2} - 1} \tag{4a}$$

$$\gamma \approx \left(q^2{\sigma_q}^2 + 3\right) q\sigma_q \tag{4b}$$

The SABR stochastic model was introduced by Hagan et al. in 2002 [5]. According to it, both stock price and volatility process are randomly driven. Under the solution of the SABR model, provided initially by its authors and refined by followers [6], the prices of European options are given through the implied Black-Scholes or Bachelier volatilities. In the following we present the vanilla call option price for the normal context with $\beta = 0$

$$V_{call} = e^{-r\Delta t}[\,(F-K)N(d_N) + \frac{\sigma_{\mathrm{N}}^{\mathrm{SABR}}\sqrt{\Delta t}}{\sqrt{2\pi}} e^{-\frac{d_N^2}{2}}] \tag{5}$$

where $d_N = (F-K)/\,(\sigma_{\mathrm{N}}^{\mathrm{SABR}}\sqrt{\Delta t})$ and the SABR implied volatility to be used in the normal model writes as

$$\sigma_{\mathrm{N}}^{\mathrm{SABR}}(K,f) = \alpha \frac{z}{x(z)}\left\{1 + \left[\frac{2-3\rho^2}{24}\right]\Delta t\right\} \tag{6}$$

with

$$z = \nu(F-K)/\alpha \tag{7}$$

$$x(z) = \log\{\mathrm{y}\} \tag{8}$$

$$y = (\sqrt{1 - 2\rho z + z^2} + z - \rho)/(1-\rho) \tag{9}$$

As discussed later on, the SABR model has the capacity to capture both the skew and smile patterns observed on the SPX and DAX option volatility surfaces.

The solution discussed above involves approximations and some recent applications are presenting extensions of it [7], [8]. Furthermore, as explained in [9], it should be mentioned that for $\nu$ functional on $q$, $\rho = \pm 1$ and $\sigma_q\sqrt{\Delta t}|q| < 3/2$, the Hagan/Oblój approximate solution of the SABR model [5], [6] is matching the SLN formulas. Separately, the SLN trajectories and prices are an exact solution of the SABR equations for $\rho = \pm 1$ [9].

## II. SPX AND DAX DATA SET ANALYSIS

For the purpose of our study we collected data on end-of-day European put and call prices for SPX and DAX index options. The data set ranges over the period 19.12.05 - 31.03.2012. In both cases we chose three price dates with an interval of one year: 18.03.10, 21.03.11 and 19.03.12. As a next step, the SLN and SABR models were fitted to the obtained data set. We discuss here the outline, while a detailed analysis will be presented in [10].

SPX: Analyzing SPX out-of-the money (OTM) option prices and their corresponding Black-Scholes volatilities for all price dates and different expiries, we observe a strong negative skew pattern, which increased from 2010 until 2012. For some price dates a mild smile pattern can be depicted. Moreover, OTM put plots reveal straight line behaviour for all price dates and all expiries. This translates into a straight volatility curve.

DAX: In this case, we observe a much stronger, dominating smile pattern comparing to the SPX data set. At the same time, the skew is less pronounced and is captured good only in the centre of the distribution, close to ATM point. Unlike the SPX OTM put plots, in this case, we distinguish a complex pattern, which cannot be analyzed without a study on bid-ask spreads.

Summary on fitting DAX and SPX prices: The central conclusion is that for expiries of 60 days or more, SLN fits accurately the SPX and DAX option prices near ATM point. The near ATM condition is defined as the price of the OTM put being no less than 10% of the ATM option price and the price of the ATM call being no less than 20% of the ATM price.

The feature mentioned above allows us to have a very precise and stable indicator of the skew parameter in option pricing without referring to the far tails. This "near-tail" approach is very practical as far tails cannot be determined accurately due to ask-bid spreads and potentially poor liquidity of deeply OTM options.

Even for pricings that can be quite well matched with the SLN model, we typically notice some deflections on the far tails. This corresponds to a deficiency of the fourth central moment of the underlying probability distribution. Such a "smile deficit" can be adjusted as a perturbation on top of the already established sigma and skew pattern. A potential practical way is to establish $\overline{\sigma} = \sigma\sqrt{\Delta T}$ γ using the SLN model fitted to near-at-the-money prices, and next implement the SABR fit to determine the full smile. The detailed iterative procedure will be discussed in a future publication [10].

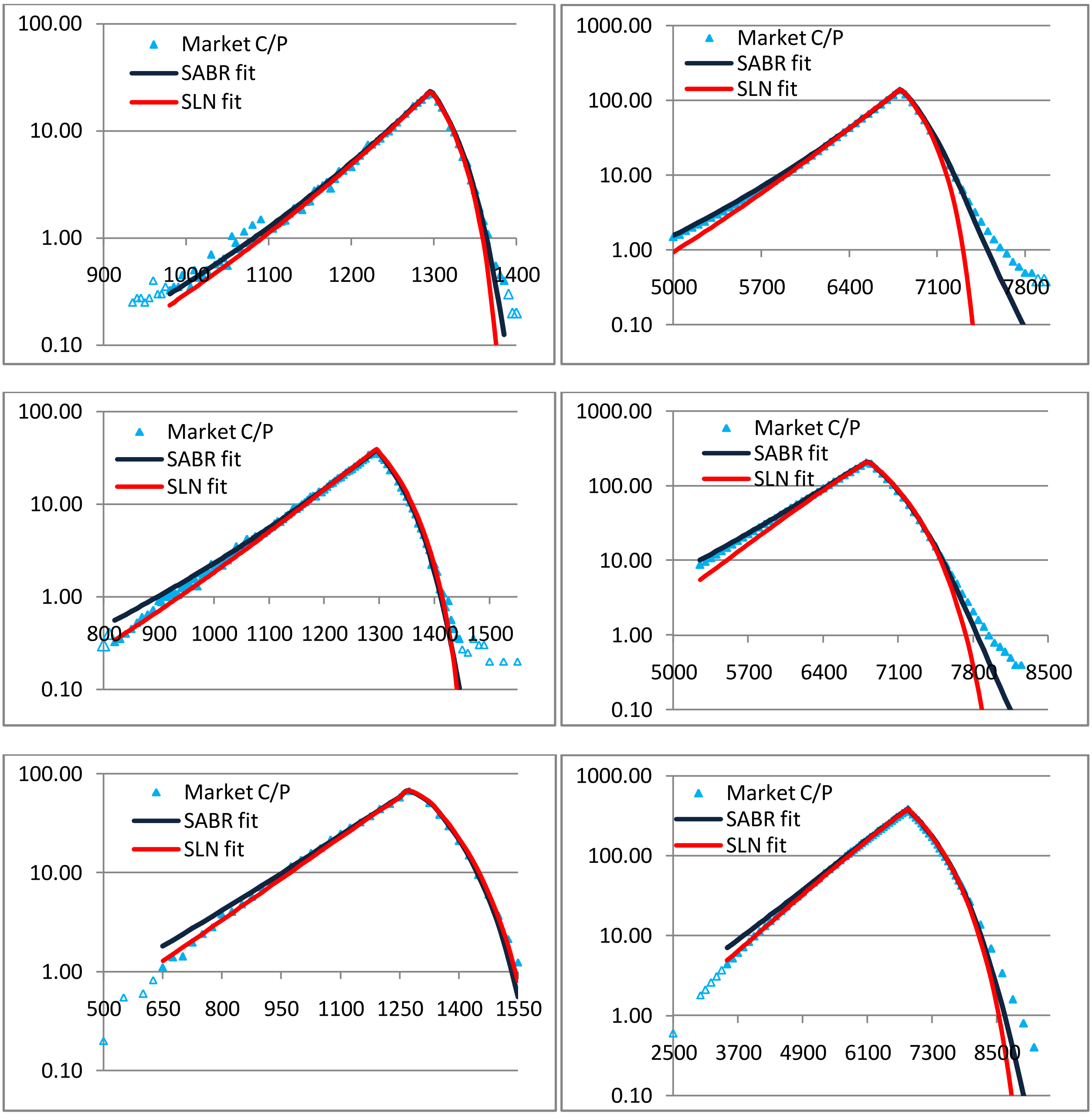

**FIG.1** Price date - 21.03.2011. SLN and SABR fit for index options with 26, 61 and 180 days to expiry. Left – SPX index option prices; right – DAX index option prices.

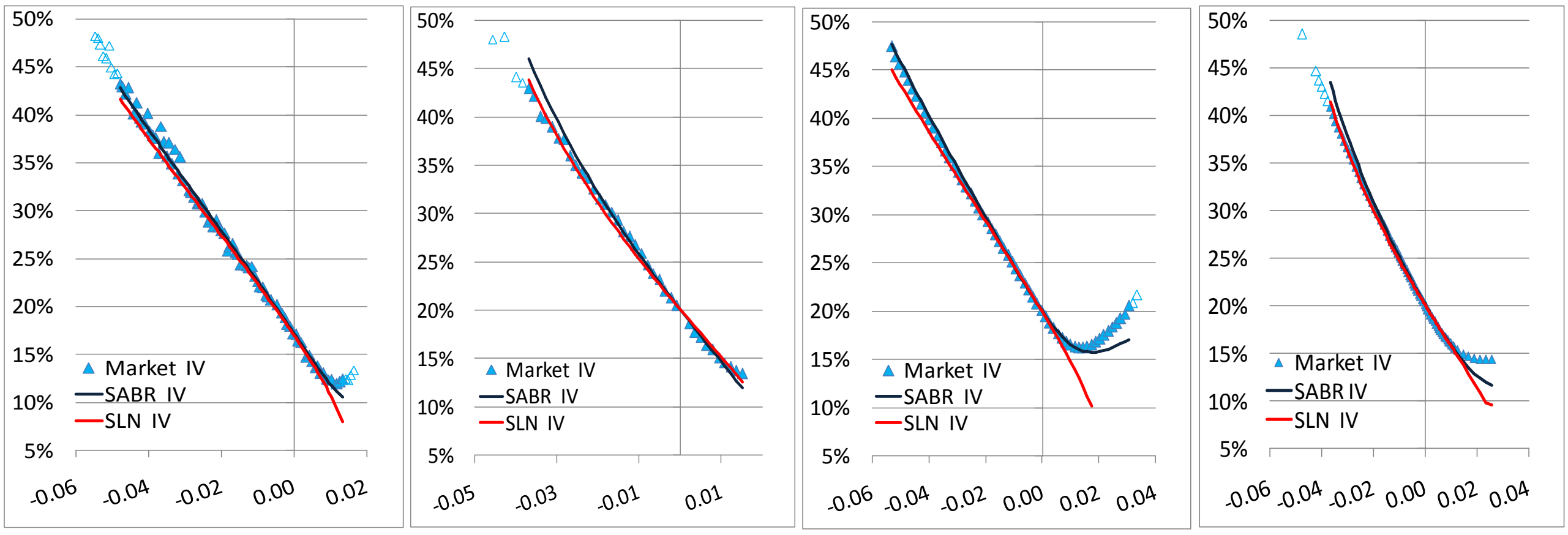

**FIG.2** Price date - 21.03.2011. SLN and SABR implied volatility (IV) fit for index options with 26 and 180 days to expiry. Left – SPX implied volatility; right – DAX implied volatility.

The SLN fit to the SPX option data set is very good for maturities of 60 days and more. The fit accuracy is high enough to state that a skew only picture, based on the SLN model, gives a quite accurate price modelling. In order to increase the accuracy and also to include the modelling of short expiries, a smile pattern needs to be added.

The SLN fit to DAX index options is poor for short term expiries, with the model failing to capture the pronounced smile pattern, but it is often accurate for long ones. Besides this, SLN provides a good approximation of the prices near ATM point. Unlike in the previous case, adjusting the smile is an indispensable condition for all expiries.

The discussed set of market option prices was next fitted with the SABR model.

For SPX options, the conclusions are:

A) Short term expiries, on the left fat tail, are priced accurately providing the option prices are no less than 1% of the ATM price.

B) For longer expiries the SABR prices of the OTM puts may show some discrepancies for strikes lower than 75% of the ATM point. In these situations, SLN may perform slightly better.

For DAX options, the conclusions fitting SABR are:

A) For mid expiries, SABR fits very well the DAX prices. Occasionally, we may detect that the SABR smile is not large enough on the OTM call side.

B) For longer expiries SABR may not price correctly the deep OTM puts and also may not catch the smile in the OTM calls.

Comparing to SLN, the SABR model involves one more parameter, which eventually leads to a better fitting capacity. However, translating individual SABR fits into a global fit across expiries typically proves to be very hard. Hence, making a global fit across all expiries may be easier using the SLN model, if the approximation is good enough.

For SPX options discussed above, SLN fitting can be interpolated in a way that both $\overline{\sigma}(\Delta T)$ and $q(\Delta T)$ are following a power law such that $\overline{\sigma}(\Delta T) = a_1\,(\Delta T)^{b_1}$ and $q(\Delta T) = a_2\,(\Delta T)^{b_2}$. Regarding $\overline{\sigma}(\Delta T)$ the power law indicator is ranging from 0.57 to 0.68 for SPX and from 0.52 to 0.60 for DAX. The $q$ factor is obeying an almost reverse power law and eventually $\chi(\Delta T) = \overline{\sigma}(\Delta T)q(\Delta T)$ has a slowly varying pattern (please note that $\chi = \sigma_q q(\Delta T)\sqrt{\Delta T}$ ). Using a polynomial fit for the Forward curve and for $\chi(\Delta \mathrm{T})$ allows us to define a simple model for the entire volatility surface with relatively few fitting parameters. We call this model a Globally fitted SLN model or GSLN.

The key is that such a modelling method gives a closed-form expression for trajectories, which makes virtually all pricing exercise straightforward. Even when Monte-Carlo is needed, say to price auto-callable products or exotic options, such a Monte-Carlo calculation is very simple. This is exactly opposite to the fits made with the SABR model, when closed form expressions for trajectories do not exist.

## III. CONCLUSIONS

SLN proves to be very efficient in modelling near-at-the-Forward prices of SPX and DAX options. Such a match allows the calculation of market skewness based on option prices, without involving higher moment calculations - hence skewness becomes a very robust quantity.

On some strongly skewed markets, such as SPX options, the SLN model can be used for a wider range of strikes. In such a case, the long left tail of OTM put options is often well matching SLN. For the same SPX options, the SABR model is giving a better and overall sufficient matching capacity.

The SLN model can be exactly mapped into the SABR model by setting the correlation parameter $\rho = \pm 1$ [9]. Eventually, by fitting option prices first to the SLN model and next to the SABR, we are coming down to a simple universal picture for modelling option prices in which the SLN is used to establish the market skew and the SABR model to evaluate the smile if necessary. This “skew first smile next” approach can be also accomplished without applying the SABR model [11].

Using solely the SLN model, when applicable, presents several advantages. The SLN model has explicit simple expressions for stochastic trajectories, which are very similar to those of the Black-Scholes model. In the following, advanced modelling is straightforward and calibration of the Monte-Carlo simulations is elementary in contrary to the very computationally demanding calibrating scenario for the SABR. Finally, it should be mentioned that the need of explicit solutions for stochastic models without odd effects associated to approximations, is recently very much discussed in the literature related to pricing of financial derivative products [12]. In this context exploiting fully solvable models like SLN and its extensions is of practical importance [11].